**Crater shape as a possible record of the impact environment of metallic bodies:**

**Effects of temperature, impact velocity and impactor density**

Ryo Ogawa[a], Akiko M. Nakamura[a], Ayako I. Suzuki[b], and Sunao Hasegawa[b]

a Department of Planetology, Graduate School of Science, Kobe University, 1-1 Rokkodai-cho, Nada-ku, Kobe 657-8501, Japan

b Institute of Space and Astronautical Science, Japan Aerospace Exploration Agency, 3-1-1 Yoshinodai, Chuo-ku, Sagamihara, Kanagawa 252-5210, Japan

Tables: 5

Figures: 12

**Send editorial correspondence to**:

Akiko M. Nakamura

Department of Planetology, Graduate School of Science, Kobe University,

1-1 Rokkodai, Nada-ku, Kobe

657-8501, Japan

e-mail: amnakamu@kobe-u.ac.jp

**Abstract**

Metallic bodies that were the cores of differentiated bodies are sources of iron meteorites and are considered to have formed early in the terrestrial planet region before migrating to the main asteroid belt. Surface temperatures and mutual collision velocities differ between the terrestrial planet region and the main asteroid belt. To investigate the dependence of crater shape on temperature, velocity and impactor density, we conducted impact experiments on room- and low-temperature iron meteorite and iron alloy targets (carbon steel SS400 and iron-nickel alloy) with velocities of 0.8–7 km s$^{-1}$. The projectiles were rock cylinders and metal spheres and cylinders. Oblique impact experiments were also conducted using stainless steel projectiles and SS400 steel targets which produced more prominent radial patterns downrange at room temperature than at low temperature. Crater diameters and depths were measured and compiled using non-dimensional parameter sets based on the $\pi$-group crater scaling relations. Two-dimensional numerical simulations were conducted using iSALE-2D code with the Johnson–Cook strength model. Both experimental and numerical results showed that the crater depth and diameter decreased with decreasing temperature, which strengthened the target, and with decreasing impact velocity. The decreasing tendency was more prominent for depth than for diameter, i.e., the depth/diameter ratio was smaller for the low temperature and low velocity conditions. The depth/diameter ratios of craters formed by rock projectiles were shallower than those of craters formed by metallic projectiles. Our results imply that the frequency distribution of the depth/diameter ratio for craters on the surface of metallic bodies may be used as a probe of the past impact environment of metallic bodies.

# 1. Introduction

Iron meteorites are composed of iron-nickel alloys and classified structurally into hexahedrites (5–6.5 wt% Ni), octahedrites (6–12 wt% Ni), and ataxites (~10 to >20 wt% Ni). Iron meteorites are also classified chemically in terms of the concentrations of trace elements, for example Ge and Ga, into several groups (e.g., Goldberg et al., 1951). Comparisons of the chemical trends within groups suggest that there are two different types: magmatic and non-magmatic (Wasson, 1985). Magmatic groups are thought to be derived from the metallic cores of differentiated bodies whereas non-magmatic groups may come from bodies that were not heated enough to form metallic cores (Haack and McCoy, 2004). The cores formed in the parent bodies of magmatic iron meteorites less than ~1.0 Myr after the formation of calcium-, aluminum-rich inclusions (CAIs) (Markowski et al., 2006; Burkhardt et al., 2008), whereas chondrule ages are about 2–4 Myr after CAIs (Scott and Sanders, 2009).

Bottke et al. (2006) suggested that the parent bodies of iron meteorites were formed in the terrestrial planet region (0.5–2 AU from the Sun), differentiated early in solar system history by the decay of $^{26}$Al, and then scattered and driven into the main asteroid belt by gravitational interactions with protoplanets. Asteroid 16 Psyche is the largest metal-rich asteroid in the main asteroid belt, measuring 232×189×279 km (Shepard et al., 2017), and might be such a remnant: it is the target of the Psyche mission (Elkins-Tanton et al., 2017). The equilibrium temperature of the iron meteorites' parent bodies decreased from about 300 K to 160 K during migration from the terrestrial planet region to the main asteroid belt. Some iron meteorites undergo a transition from ductile to brittle behavior as the temperature decreases (Johnson and Remo, 1974). Octahedrite iron meteorites (Gibeon, El Sampal, and Arispe) showed brittle behavior at <200 K in previous impact cratering experiments (Matsui and Schultz, 1984). Cooled steel (SS400 carbon steel) targets were shown to produce finer fragments compared to those at room temperature (Katsura et al., 2014). Impact cratering experiments on iron meteorites and foundry-cast ingots showed that cratering efficiency was lower for lower temperatures (Marchi et al., 2020). Accordingly, it is expected that the shape of impact craters on iron meteorites' parent bodies is dependent upon temperature and, therefore, heliocentric distance. Moreover, the most probable collision velocity in the main asteroid belt is 4.4 km/s (Bottke et al., 1994), however, it could have been higher in the terrestrial planet region early in the solar system (Bottke et al., 2006). Such a velocity difference may also cause a difference in crater shape on metallic bodies.

Crater formation is affected by both the material strength of, and gravity at, the surface of the target (Holsapple, 1993). The relative influence of strength and gravity on cratering efficiency can be assessed by the ratio $S = \frac{Y}{\rho g L}$, where $Y$, $\rho$, $g$ and $L$ are the strength, density, gravitational acceleration and impactor diameter, respectively. Large values of $S$ $(> 10)$ imply strength-dominated cratering (Collins et al., 2011). If we assume a spherical metallic body of 200 km in diameter with bulk density of 7.8 g cm$^{-3}$ and static strength of 500 MPa, $L \approx 30$ km corresponds to $S \approx 10$. A previous impact cratering experiment showed that an $L = 1$ mm stainless steel projectile forms a crater of 3.3 mm diameter on a stainless steel target by vertical impact at 5 km s$^{-1}$ at room temperature (Burchell and Mackay, 1998). Therefore, it is expected that most craters on 200 km metallic bodies form in the strength regime, if the surface regolith is not considered and excepting the largest possible craters.

In this paper we describe laboratory impact cratering experiments on, and numerical simulations of, Gibeon iron meteorite, SS400 and iron-nickel alloy targets at room and low temperatures. Velocities of 0.8–7.0 km s$^{-1}$ and 1.0–25 km s$^{-1}$ were used in the laboratory experiments and the numerical simulations, respectively. We focused on the effects of temperature and impact velocity on crater shape. We also discuss the effect of projectile material on the depth/diameter ratio. Modification of the surface composition and texture of the Gibeon target due to impact was reported in a separate paper (Libourel, et al., 2019). Although a slight size-scaling effect on crater diameter on aluminum targets has been reported (Walker et al., 2020), here we simply extrapolate the results of this study to asteroid scale. Possible detection of the impact environment, such as temperature and impactor population in terms of impact velocity and density of impactor, are then discussed, based on the combined results of the laboratory experiments and the numerical simulations.

## 2. Experiments

We performed impact cratering experiments on room- and low-temperature metallic targets using a two-stage hydrogen-gas gun installed at the Institute of Space and Astronautical Science (ISAS) and a vertical powder gun at Kobe University. Cylindrical projectiles of dunite or basalt were used to simulate rocky impactors. These were cut from leftover samples from previous impact experiments (Katsura et al., 2014; Nakamura and Fujiwara, 1991). The bulk densities of the dunite and basalt were 3.2 and 2.7 g cm$^{-3}$, respectively. Projectiles of stainless steel (SUS304; 7.93 g cm$^{-3}$) were used to simulate metallic projectiles. Spherical projectiles of copper (8.94 g cm$^{-3}$), titanium (4.5 g cm$^{-3}$), cemented tungsten carbide (WC-6%Co; 15 g cm$^{-3}$), aluminum (2.7 g cm$^{-3}$) or alumina (3.95 g cm$^{-3}$) were used to examine the effect of projectile density on crater shape. The targets were blocks of Gibeon iron meteorite (7.9 g cm$^{-3}$), steel (SS400; 7.85 g cm$^{-3}$) or iron-nickel alloy (8.2 g cm$^{-3}$). The number of Gibeon blocks available was limited, therefore we used the SS400 and iron-nickel alloy blocks as replacements. Gibeon iron meteorite is composed of an iron-nickel alloy that includes 8 wt% Ni (Buchwald, 1969); it undergoes a ductile-brittle transition as the temperature is lowered. Its transition temperature increases with strain rate and is expected to be 200 ± 50 K under impact conditions (Johnson and Remo, 1974). SS400 has a tensile strength of 400–510 MPa. The ductile-brittle transition of SS400 in Charpy impact tests occurs at a similar temperature to that of the Gibeon meteorite, at about 230 K (Arai et al., 2010).

After impact, the surface profile of the target was measured using a laser profiler installed at ISAS as described previously (Suzuki et al., 2012; 2021). In the following, the setups of the experiments with the two-stage hydrogen-gas gun and the powder gun are described.

### 2.1. Two-stage hydrogen-gas gun experiments

The projectiles used were dunite or basalt cylinders with a diameter of 3 mm and height of 2 mm, or stainless steel, copper, WC, titanium, aluminum or alumina spheres with a diameter of 1 or 3 mm. They were launched horizontally with velocities from 1.8 to 7.0 km s$^{-1}$ using a sabot (Kawai et al., 2010). The projectile speed was determined from the time interval for flight between two laser beams. The error of velocity estimation is 1% or less. Table 1 shows the experimental conditions for these two-stage hydrogen-gas gun shots. The targets were blocks of Gibeon or cubes of SS400, with sides 40 mm or longer, or Ni-31.5 wt% and Ni-42 wt% iron-nickel alloy with 50 mm sides. A hole with a diameter of 3 mm and depth of 10 mm was drilled for a rod-shaped temperature sensor for the cube targets. We used targets at room and low temperatures (nominally below 150 K). The targets for the low-temperature shots, including Gibeon, were cooled by liquid nitrogen to 77 K for more than 10 minutes before impact, then put on a cooled SS400 cube with 70-mm sides to keep the temperature low in the experimental chamber, in which the ambient pressure was rapidly reduced to below 0.5–5.0 Pa. The temperature was monitored during the pumping-down phase before impact and did not significantly exceed 150 K. The results of four shots of rock projectiles on Gibeon targets, which showed a resulting coat of projectile material on the surface of the craters, have been published as a separate paper (Libourel et al., 2019).

Nine oblique shots were conducted with incident angles, $\theta$, from the surface to the projectile trajectory of 10–45 degrees. The targets were SS400 rectangular parallelepiped blocks measuring 10×5×5 cm and the projectiles were stainless steel spheres. The long axis of the block was placed in the ballistic direction of the projectile. Details on how to set up the target can be found in Suzuki et al. (2021). Table 1 shows the experimental conditions of these oblique impacts.

**2.2. Powder gun experiments**

The projectiles used were stainless steel cylinders with a diameter of 15 mm and height of 15 mm, launched vertically with velocities of 0.8–1.3 km/s. The impact velocity was determined by high-speed imaging of the projectiles. Because the number of images available was limited, the error in velocity estimates is about 3%. Table 2 shows the experimental conditions of the powder-gun shots. Targets were SS400 cubes with sides of 70 mm or more. The target and a spare block were put into liquid nitrogen before each shot. Then the target block was set in the experimental chamber. The ambient pressure in the chamber was about $10^4$ Pa, but the time between setting the target and the impact was much shorter than in the two-stage hydrogen-gas gun experiments. We monitored the temperature of the spare block left outside the chamber, instead of direct temperature measurement of the target inside it.

## 3. Numerical simulation

We carried out numerical simulations of vertical impact using the two-dimensional version of the iSALE shock physics code, i.e., iSALE-Chicxulub version (iSALE-2D; Wünnemann et al., 2006) especially for hypervelocity impacts (>7.0 km s$^{-1}$), which cannot be conducted via two-stage hydrogen-gas gun experiments. iSALE is an extension of the SALE hydrocode (Amsden et al., 1980). To simulate hypervelocity impact in solid materials, SALE was modified to include an elastoplastic constitutive model, fragmentation models, various equations of state, and multiple materials (Melosh et al., 1992; Ivanov et al., 1997). More recent improvements include a modified strength model (Collins et al., 2004) and a porosity compaction model (Wünnemann et al., 2006).

We used the ANEOS model for the equation of state of iron (Thompson and Lausen, 1972) for the Gibeon iron meteorite target, the SS400 target and the SUS304 projectile and the Tillotson equation of state of copper for the copper projectile. We used a strength model based on the Johnson–Cook strength model (JNCK; Johnson and Cook, 1983, 1985) for the Gibeon, SS400,

SUS304 and copper. JNCK is one of several strength models available in iSALE-2D. Yield strength, *Y*, in this model is defined as:

$$Y = (A + B\bar{\varepsilon}_p^N)\{1 + Cln(\dot{\varepsilon}^*)\}(1 - T^{*M}) \qquad (1)$$

where $\bar{\varepsilon}_p$ is the equivalent plastic strain and $\dot{\varepsilon}^*$ is the non-dimensional plastic strain rate normalized by the reference strain rate $\dot{\varepsilon}_0 = 1\ s^{-1}$. The non-dimensional temperature *T*\* is defined as:

$$T^* = \frac{T - T_{ref}}{T_m - T_{ref}} \qquad (2)$$

where *T* is temperature, $T_{ref}$ is reference temperature, and $T_m$ is the melting point of the material. *A, B, N, C,* and *M* are constants (here after JNCK parameters), where *A* is *Y* when $\bar{\varepsilon}_p$ = 0 and $\dot{\varepsilon}^*$ = 1. JNCK parameters have previously been experimentally determined for Armco iron, oxygen-free high thermal conductivity (OFHC) copper, and other metals (Johnson and Cook, 1985). We estimated the JNCK parameters based on literature information on octahedrite iron meteorite and used these for the Gibeon iron meteorite. We also estimated the parameters for SS400 and SUS304 as described below and shown in Table 3.

### 3.1. JNCK parameters

3.1.1. Parameters for iron meteorite

We determined JNCK parameters *A, B, N, C,* and *M* for iron meteorites based on the true

stress−true strain ($\sigma - \varepsilon$) curves of an octahedrite, the Henbury meteorite (7.5 wt% Ni) (Furnish et al., 1994). In the JNCK model, the dependencies on strain, strain rate, and temperature are independent, although the measurement results show that they are not. In addition, instead of setting $T_{ref}$ = 298 K, we set $T_{ref}$ = 77 K to use it to simulate a low temperature target. As discussed below in this section and shown later in Figure 9, the determined parameter set shown in Table 3 reproduced only part of the measurement behavior of the stress-strain curves of the meteorite.

First, $C$ was determined assuming that the ratio of the yield strengths $\sigma_Y$ on the $\sigma - \varepsilon$ curves, when $\dot{\varepsilon} = 0.001$ and $4000\ s^{-1}$ at 298 K was ~1.5. Using the following relationship, we derived C = 0.027:

$$\frac{\sigma_Y(298\ \text{K}:\ \dot{\varepsilon} = 4000)}{\sigma_Y(298\ \text{K}:\ \dot{\varepsilon} = 0.001)} = \frac{Y(298\ \text{K}:\ \dot{\varepsilon} = 4000)}{Y(298\ \text{K}:\ \dot{\varepsilon} = 0.001)} = \frac{1 + C\ln(4000)}{1 + C\ln(0.001)} \sim 1.5. \qquad (3)$$

Note that the value of $C$=0.027 gives $\frac{Y(\dot{\varepsilon}=3000)}{Y(\dot{\varepsilon}=0.001)} = \frac{1+C\ln(3000)}{1+C\ln(0.001)} = 1.495$; however, the ratio of the measured yield strength at 77 K (when $\dot{\varepsilon} = 0.001$ and $3000\ s^{-1}$) is approximately unity and is smaller than 1.5; in other words, parameter $C$ is not a constant, but is dependent on temperature.

Based on Equations 1 and 2, the relation between $A$ and $Y$ at 77 K, $Y(77K: \dot{\varepsilon})$, is written as:

$$A = \frac{Y(77K:\ \dot{\varepsilon})}{1 + C\ln(\dot{\varepsilon})}. \qquad (4)$$

Inserting the value of $\sigma_Y$ of the Henbury meteorite when $\dot{\varepsilon} = 0.001$, ~550 MPa, at 77K (Furnish et al., 1994) into $Y$ in Equation 4 gives $A \approx 680$ MPa. The parameters $B$ and $N$ are determined from the plastic part of the stress-strain curve, that is, the strain hardening ($\Delta\sigma = \sigma - \sigma_y$)–plastic strain ($\varepsilon_p = \varepsilon - \varepsilon_y$, where $\varepsilon_y$ is the yield strain) curve. We assumed a power law relationship between the strain hardening and plastic strain, such that:

$$\Delta\sigma = B\varepsilon_p^N \tag{5}$$

and obtained $B = 1015$ MPa and $N = 0.53$. Finally, the thermal parameter $M$ was determined by the following equation:

$$M \sim \frac{\log\left\{1 - \frac{\sigma_Y(298\ \text{K}:\ \dot{\varepsilon})}{\sigma_Y(77\ \text{K}:\ \dot{\varepsilon})}\right\}}{\log\left\{\frac{298 - 77}{T_m - 77}\right\}}, \tag{6}$$

where $\sigma_Y(298\ \text{K}:\ \dot{\varepsilon})$ and $\sigma_Y(77\ \text{K}:\ \dot{\varepsilon})$ are the yield strengths for the same strain rate at each temperature, respectively. Note that the experiment data show that the ratio $\frac{\sigma_Y(298\ \text{K}:\ \dot{\varepsilon})}{\sigma_Y(77\ \text{K}:\ \dot{\varepsilon})}$ is not constant but varies with strain rate. As shown in Table 3, we assumed a lower $T_m$ (= 1783 K) than that used for iron (1811 K) corresponding to the lower value for Ni than for Fe (Swartzendruber et al., 1991). We used a value of 0.52 for $M$.

3.1.2. Parameters for SS400 and SUS304

We determined the JNCK parameters $A$, $B$, $N$, $C$ and $M$ with $T_{ref}$ fixed at 298 K for SS400 based on data obtained at room temperature. We first derived the parameters $A$ and $C$ using the data for $Y$ of SS400 at a different strain rate (Shimada et al., 2012) by assuming $T = T_{ref}$:

$$Y(\dot{\varepsilon}) = A\{1 + C\ln(\dot{\varepsilon})\}. \tag{7}$$

We obtained $A = 360$ MPa and $C = 0.045$. Second, we obtained $B = 555$ MPa and $N = 0.49$ according to Equation 5 using the plastic part of the stress-strain curve of SS400 at 298 K and $\dot{\varepsilon} = 2.5 \times 10^{-5}$ s$^{-1}$ (Itoh et al., 2001). Finally, the temperature dependence of $Y$ of SS400 obtained at $\dot{\varepsilon} = 0.0025$ (JIS G3101 (2004)) was used to determine $M$ according to:

$$Y(\dot{\varepsilon} = 0.0025 : T) = 360\{1 + 0.045 \ln(0.0025)\}(1 - T^{*M}). \tag{8}$$

Here we assumed $T_m = 1811$ K. The value of $M$ was derived as 0.602. The same procedure was followed to obtain the parameters of SUS304 based on data from the literature (Li et al., 2013; JIS G 4304 (2005)).

### 3.2. Simulation setup

We defined the geometry of the mesh as having cylindrical symmetry. The mesh had a high-resolution zone close to the impact site, which was 150–200 and 250–350 cells in size in the x- and y-directions, respectively, and extension zones below, above and to the right of the high-resolution zone to mitigate the effects of reflections of the shock wave from the mesh boundaries. We chose a spatial resolution of 20 cells per projectile radius based on previous studies (Pierazzo et al., 2008; Davison et al., 2011). Simulations were performed with impact velocities of 1–25 km s$^{-1}$, including conditions similar to the Gibeon target experiments (Gd1–3 and Gs1–4) and three shots of the SS400 target experiments (Sc1–3) as shown in Table 4. The material properties of dunite were used for the projectile in simulations Gd1–3, while those of SUS304 were used for the projectile in Gs1–4. We used the ANEOS equation-of-state of dunite ($\rho_p$=3.32 g cm$^{-3}$) (Benz et al., 1989), a strength model of rock (Collins et al., 2004), and a damage model (Ivanov et al., 1997) according to the simulation of a basin forming impact on asteroid 4 Vesta (Bowling et al., 2014) for the dunite projectile. The initial temperature of the target was set at 298 K and 132 K for the room- and low-temperature conditions, respectively. The simulation ended when the crater rim stopped growing.

## 4. Results

### 4.1. Measurement results

All of the experimental craters have overturned and partly disrupted crater rims. Figures 1a–d shows optical images of the Gibeon targets after the shots. For impacts of dunite projectiles (Gd1–3), particularly the high velocity shot (Gd2), fractures can be seen near the crater, as shown in Figure 1a–c. Scanning electron microscopy images of these craters have been reported in a previous study (Libourel et al., 2019). No obvious temperature dependence of the structure of the crater rim or the fractures around the crater was observed in the images of the Gibeon targets in this study, in contrast to a previous study (Matsui and Schultz, 1984). Figure 1e–h shows optical images of the craters on Gibeon targets formed by stainless steel projectiles.

The Gibeon blocks in this study are small compared to the crater size. For brittle targets, the tensile strength is much lower than compressive or shear strength; therefore, when the target is small, the rarefaction waves generated at the free surface affect the spall diameter and total crater volume, with the topographic profile of the central pit being unaffected (Suzuki et al., 2018). Metal targets are less susceptible to rarefaction waves than are brittle targets because of their high tensile strength. The crater shape has a roughly circular symmetry, although the rim shape is highly irregular and seems to be affected by local discontinuities (fractures) more than the boundaries of the targets. The target looked intact before the Gs2 shot, as shown in Figure 2a, however, the interiors of the targets were oxidized, which may have affected the fracture patterns. Figure 2b shows the target for which a shot produced the largest fracture; brown zones appeared along the fractured surfaces.

Figure 3 compares the craters formed on SS400 targets by oblique impacts. The crater became elongated along the velocity component and shallower as $\theta$ became smaller. At the shallowest $\theta$ (= 10°), the cavity consisted of multiple depressions along a line in the direction of the velocity component, as shown in a previous study (Burchell and Mackay, 1998). The downrange damage is due to a decapitated projectile (Burchell and Mackay, 1998; Davison et al., 2011; Suzuki et al., 2021). Differences among the appearances of crater cavities formed at different temperatures are not clear as shown in Figure 6 below, although the patterns extending downrange of the crater are more obvious for those at room temperature. It has been reported that the target material formed into metal beads from a few nanometers to tens of micrometers in size mixed with melted dunite projectile material in the crater of Gd2 (the room temperature, 7 km/s shot) (Libourel et al., 2019). Ejecta from the normal impact of silicate projectiles into steel targets have been found to contain metal beads (Ganino et al., 2019). Such metal beads may have formed the splatter pattern found in the highly oblique shots in this study. The ejecta pattern of a highly oblique impact provides a clue regarding the impact conditions, not only the impact angle but also the target temperature and impact velocity for a metallic surface.

Figure 4 shows the definitions of the crater diameter, $D$, crater depth, $d$, and height of the crater rim, $h$, in this study shown in Tables 1 and 2. For oblique impact experiments, the length of the crater along the velocity vector $D_{length}$ and the width perpendicular to this $D_{width}$ were measured. For normal impacts, the diameter uncertainty was typically a few percent (considering the variation in each measurement direction) and the depth uncertainty was $\leq 0.1$ mm. $h$ was measured for four points at either end of two roughly orthogonal profiles taken through the crater formed by a normal impact shown in Table 1. Due to the variation in rim height along the periphery, the values of rim height are affected by the choice of measurement directions. Therefore, the values of rim height shown in Tables 1 and 2 contain large uncertainty.

Figures 5a, b shows $D$ versus impact energy and $d$ versus impact energy. Fig. 5c shows the depth/diameter ratio, $\frac{d}{D}$, versus the normal component of the impact velocity $U$, i.e. $U\sin\theta$. Here, we plot $\frac{d}{D_{mean}}$, where $D_{mean} = \sqrt{D_{length}D_{width}}$, for an oblique impact. We discuss the dependence on the impact angle below. $D$ is only weakly dependent on the projectile material and temperature, in contrast to $d$. Depth tends to be shallower at low temperature, therefore, $\frac{d}{D}$ values of room-temperature targets tend to be larger than those at low temperature, although the ratio also depends on other factors such as projectile material and impact velocity, as shown in a previous study of metallic targets (Burchell and Mackay, 1998). For example, craters formed by a WC-6%Co projectile have $\frac{d}{D}$ values a few times larger than those of other projectiles. The craters of the powder-gun shots ($\leq$1.3 km s$^{-1}$) and those formed by highly oblique incidences had smaller $\frac{d}{D}$. We compare the results of $D$ and $d$ from our simulations with those of our experiments in Figure 5d, e. The differences between $D$ and $d$ from the experiments and the simulations were within 5% and 17%, respectively.

Figure 6 shows the data from oblique impact experiments on SS400 targets performed with an impact velocity of 5 km s$^{-1}$. The ratio $\frac{D_{length}}{D_{width}}$ is defined as the ellipticity (Bottke et al., 1994). It can be seen in Figure 6a that the ellipticity increases as $\theta$ decreases. The ellipticities we measured are lower than those of the craters formed by stainless steel projectiles with an impact velocity of 5 km s$^{-1}$ on stainless steel targets (Burchell and Mackay, 1998). A numerical simulation of oblique impacts on metal targets showed that ellipticity is lower for targets with lower strength

(Davison et al., 2011). The strength of SS400 is lower than that of stainless steel, therefore, the lower ellipticity of SS400 craters is consistent with the numerical study. No temperature effect is evident in our data for θ ≥ 20°, and the large difference in the results for $\theta = 10°$ might have been due to the stochastic nature of the breakup of the projectile.

Figure 6b shows the mean diameter, $D_{mean}(\theta)$, and depth, $d(\theta)$, of the crater. These are normalized using the average diameters (11.0 ± 0.6 and 11.0 ± 0.3 mm for room and low temperatures, respectively) and depths (6.7 ± 0.3 and 6.0 ± 0.2 mm, respectively) of the craters formed in vertical impacts with a velocity of about 5 km s$^{-1}$ (4.98 ± 0.08 and 4.96 ± 0.20 km s$^{-1}$, respectively). The error bars displayed correspond to the scatter of the values for vertical incidence. In Figure 6b, the previous results for a stainless steel projectile and target (Burchell and Mackay, 1998) are also shown. Our present results for the SS400 target are similar to those previous data. No temperature dependence was observed between the results for normalized diameter and depth, especially for normalized depth, i.e., the ratio of crater depth at an oblique incidence to normal incidence. In other words, the effects of impact angle were indistinguishable between low temperature and room temperature.

Power-law relationships were obtained using both the present and previous data; these are shown in Figure 6b by solid curves:

$$\frac{D_{mean}(\theta)}{D} = sin^{0.316\pm0.022}\theta, \tag{9a}$$

and

$$\frac{d(\theta)}{d} = sin^{1.076\pm0.029}\theta. \tag{9b}$$

A previous study of numerical simulations of oblique impacts on a metal target with a strength of 200 MPa reported $D_{length}(\theta) \propto sin^{0.22}\theta$ and $D_{width}(\theta) \propto sin^{0.46}\theta$ (Davison et al., 2011), meaning $D_{mean}(\theta) \propto sin^{0.34}\theta$. The work of Davison et al. (2011) also showed $d(\theta) \propto sin\theta$, which is consistent with Equation 9b. However, Equation 9a underestimates the mean crater diameter at $\theta \geq 45°$. To reproduce the diameter data at $\theta \geq 45°$ better, a modified empirical relationship is obtained:

$$\frac{D_{mean}(\theta)}{D} = (sin\theta)^{0.677\pm0.096}[\exp\{(0.67 \pm 0.16)(1 - sin\theta)\}]. \tag{10}$$

Note that Equations 9 and 10 were based on data from experiments conducted at about 5 km s$^{-1}$, while previous experiments showed that the dependence of crater depth, length, and width on impact angle varies with $U$ and the combination of projectile and target (Burchell and Mackay, 1998).

According to Equations 9b and 10, $\frac{d}{D}$ can be represented as:

$$\frac{d(\theta)}{D_{mean}(\theta)} = (sin\theta)^{0.40} \exp\{0.67(sin\theta - 1)\}\left(\frac{d}{D}\right). \tag{11}$$

Figures 7 shows $h$ versus $D$. The data is very scattered, however, $h$ may tend to be slightly larger for craters formed at room temperature than those at low temperature. This is consistent with the tendency toward deeper craters formed at room temperature shown in Figure 5b, c. Note that a coating of the projectile material on the crater floor and wall was visible in most of our experiments on metal projectiles, as shown in the cross-sectional images of craters in Figure 8a–c. The thickness of the coating was thicker on the bottom of craters than on the wall. It was shown that the material of rock projectiles coated the interior of the craters and modified the reflectance spectra (Libourel et al., 2019). Figure 8a–c also shows that the craters formed at lower velocity (Figure 8a, b) have shallower cavities and smoother walls, indicative of plastic deformation of the target with the projectile becoming embedded, while the crater formed at higher velocity (Figure 8c) has a rugged surface and bowl shape as a result of intense excavation and the ejection of target material. In another study, a similar transition of the crater shape and wall roughness with impact velocity was observed for aluminum targets (Nishida et al., 2012).

4.2. π-group scaling

We compared $D$ and $d$ quantitatively using π-group scaling (according to Equation 6 in Holsapple and Schmidt, 1982),

$$\pi_D = \left(\frac{\rho_t}{m}\right)^{\frac{1}{3}} D, \quad \pi_d = \left(\frac{\rho_t}{m}\right)^{\frac{1}{3}} d, \quad \pi_V = \frac{\rho_t V}{m}, \quad \pi_3 = \frac{Y_s}{\rho_p U^2}, \quad \pi_4 = \frac{\rho_t}{\rho_p}, \quad (12)$$

where $V$ is the crater volume, $Y_s$ is the strength of the target material, $U$ is the impact velocity, and $\rho_t$ and $\rho_p$ are the target and projectile densities, respectively. $\pi_D$, $\pi_d$ and $\pi_V$ are considered to depend on $\pi_3$ and $\pi_4$ in the strength-dominated regime (e.g., Holsapple and Schmidt, 1982) as,

$$\pi_D = K_D \pi_3^{-\alpha_D} \pi_4^{\beta_D}, \quad \pi_d = K_d \pi_3^{-\alpha_d} \pi_4^{\beta_d}, \quad \pi_V = K_V \pi_3^{-\alpha_V} \pi_4^{\beta_V}. \tag{13}$$

where $K_D$, $K_d$, $K_V$, $\alpha_D$, $\alpha_d$, $\alpha_V$, $\beta_D$, $\beta_d$, and $\beta_V$ are constants. Previous results of laboratory cratering experiments on geologic materials were often compiled using the static strength such as tensile strength of the target (e.g., Housen and Holsapple, 2011; Suzuki et al., 2012; Nakamura, 2017). Figure 9 shows the temperature dependence of the tensile stress and the yield strength ($Y$) of the Henbury iron meteorite (Furnish et al., 1994), and $Y$ of the Gibeon iron meteorite (Gordon, 1970). We show $Y$ of the JNCK model calculated from the parameters of an iron meteorite determined in section 3.1.1. We assumed that the temperature dependence of tensile stress is the same as that of yield strength, and obtained:

$$Y_t = Y_{t\_ref}(1 - T^{*M}), \tag{14}$$

where $Y_{t\_ref} = 890$ MPa, based on the Henbury data in Figure 9. We calculated the value of $Y_s = Y_t$ for π-group scaling according to Equation 14 for the low-temperature Gibeon target.

To obtain the relationships in Equation 13 for the Gibeon results, we conducted least squares fits to the data of this study. We also fitted power-law relationships between $\pi_D$ and $\pi_d$ versus $\pi_3$ for the dataset of a stainless steel projectile with an SS400 target to obtain the indices of $\alpha_D$ and $\alpha_d$ for room and low temperature, respectively. In this procedure, we assumed the room temperature $Y_s$ (= 480 MPa) for the SS400 target, irrespective of the temperature condition, which should not affect the values of $\alpha_D$ and $\alpha_d$. Note that the value of $\pi_4$ is 0.990 and is approximately unity for this projectile-target combination, that is $\pi_4^{\beta_D} \sim \pi_4^{\beta_d} \sim 1$. The results are shown in Table 5, and collectively show that $\alpha_D \sim 0.27$ and $\alpha_d \sim 0.5$ for iron material. The value of $\alpha_d \sim 0.5$ shows that $d$ is proportional to $U$ with a power index of 1, which is close to the power index of $\sin\theta$ (1.08) shown by Equation 9b. Accordingly, we could use a modified $\pi_3' = \frac{Y_s}{\rho_p (U\sin\theta)^2}$ for $d$. The replacement of $U$ by the vertical component of the velocity vector was suggested in previous studies (e.g., Chapman and Makinnon, 1986; Davison et al., 2011).

If the crater shape does not change with the impact velocity and material parameters, it is expected that $\alpha_V = 2\alpha_D + \alpha_d$ and $\beta_V = 2\beta_D + \beta_d$[1]. In this study, $2\alpha_D + \alpha_d = 1.1 \pm 0.3$ and $2\beta_D + \beta_d = 1.0 \pm 0.6$, while $\alpha_V = 0.709 \pm 0.030$ and $\beta_V = 0.523 \pm 0.037$ were obtained in a previous study of craters on metal targets (Holsapple and Schmidt, 1982). As shown in Figure 8a–c, the crater shape is not the same but varies with impact conditions, therefore, the discrepancies between the power-law indices are natural. Note that according to an empirical relationship for penetration depth into an infinite target for $\frac{\rho_p}{\rho_t} < 1.5$ (Christiansen, 1992, and references therein):

$$d \propto (2a)^{\frac{19}{18}} H^{-0.25} \left(\frac{\rho_p}{\rho_t}\right)^{0.5} \left(\frac{U}{C}\right)^{\frac{2}{3}}, \quad (15)$$

where $a$, $H$, and $C$ denote the projectile radius, Brinell hardness of the target, and sound speed in the target, respectively. From this, $\beta_d = 0.17$ is inferred and is consistent with the value obtained for the Gibeon target data.

---

[1] The relation between crater volume and diameter and depth can be represented as: $V = K_s D^2 d$, where $K_s$ is a shape factor that can depend on impact velocity and material parameters such as projectile and target density and porosity, and target strength. Substituting Equations 13 and 14 into $V = K_s D^2 d$, we obtain: $K_V \pi_3^{-\alpha_V} \pi_4^{\beta_V} = K_s K_D^2 K_d \pi_3^{-(2\alpha_D + \alpha_d)} \pi_4^{2\beta_D + \beta_d}$.

Figure 10a and b show the experimental and numerical results of $\pi_D/\pi^{\beta_D}$ and $\pi_d/\pi^{\beta_d}$, respectively, for the Gibeon target of this study, the results of the SS400 target conducted at room temperature (except for the WC projectile), and the results of our numerical simulations of the impacts on SS400 by a copper projectile. Almost all of the experimental results in Figure 10a, b follow the empirical relationships obtained for the Gibeon target data, irrespective of the temperature condition. One outlier among the experimental results is the shot of a basalt projectile into an SS400 target. For this, it is possible that the projectile broke during acceleration in the gun muzzle and, as a result, the $D$ and $d$ became smaller. $D$ for numerical simulations of low $U$ (1.0 and 1.5 km s$^{-1}$, that is, $\pi_3 > 10^{-1}$) for rock projectiles are larger than the extrapolation of the empirical relationship because the diameter of the crater cannot be smaller than the diameter of the cylindrical projectile, while the values of $d$ of the two lowest $U$ cases follow the empirical relationship. As for $d$, the slope of the $\pi_d - \pi_3$ relationship at high $U$ ($\pi_3 \leq 10^{-3}$) appears to be shallower than for low $U$.

Let us discuss the velocity dependence of $D$ and $d$. The diameter and depth are proportional to $U$, to the power of ~0.5 and ~1, which means that $\frac{d}{D}$ is proportional to $U$, to the power of about 0.5: the higher the value of $U$, the deeper the crater. According to Equations 12 and 13, we obtain:

$$\frac{d}{D} = \frac{\pi_d}{\pi_D} = \frac{K_d}{K_D} \pi_3^{-(\alpha_d - \alpha_D)} \pi_4^{(\beta_d - \beta_D)}. \tag{16}$$

Substituting the fitting results of Gibeon data shown in Table 5, we obtain:

$$\frac{d}{D} = 0.11 \pi_3^{-0.28} \pi_4^{-0.11} \propto \rho_p^{0.39} \rho_t^{-0.11} U^{0.56} Y_s^{-0.28}. \tag{17}$$

On the other hand, previous impact experiments on sedimentary rock targets gave a set of similar exponents of $\pi_3$ for $\pi_D$ and $\pi_d$: $-0.22 \pm 0.02$ and $-0.25 \pm 0.02$, respectively (Suzuki et al., 2012). Similarly, results for igneous rock targets showed that $D$ and $d$ have similar power-law dependences on $U$, i.e., 0.740 and 0.714, respectively (Gault, 1973). These indicate that the $\frac{d}{D}$ values of craters on rock targets are almost insensitive to $U$. By contrast, the $\frac{d}{D}$ values of craters on metallic bodies are indicative of the impact velocity, as well as $\rho_p$, and $Y_s$.

Figure 11 shows the $\frac{d}{D}$ values of craters formed on Gibeon targets at room and low temperature and SS400 targets at room temperature by vertical impact in the laboratory versus $\pi_3$. The results of numerical simulation for a Gibeon target and dunite projectile at low temperature with impact velocities of 8, 14, and 20 km/s and the experimental data of a previous study of a quartz projectile striking a Gibeon target (Marchi et al., 2020) are also shown. The data of rock projectiles in this study are lower than the dotted curve, which shows $\frac{d}{D} = 0.11\pi_3^{-0.28}$. Although the data for higher density projectiles tend to locate above the data for rock projectiles, the data for a quartz projectile (density: 2.2 g cm$^{-3}$) are also above the data for rock projectiles, which indicates more complex dependence of $\frac{d}{D}$ on material properties than the power law of projectile density.

We obtained an empirical relationship that defines the lower limit of $\frac{d}{D}$, shown by the dashed curve in Figure 11, as:

$$\frac{d}{D} = 0.16\pi_3^{-0.16}, \tag{18}$$

using the experimental results for a Gibeon target and rock projectile (four data points) and the numerical results shown in Figure 11 (three data points).

## 5. Crater depth/diameter ratio distribution on a metallic body

This study shows that $d$ depends more on $Y_S$ and $U$ than $D$ does in cratering on metal targets, including Gibeon targets. Accordingly, it is expected that $\frac{d}{D}$ for a metallic body depends on $Y_S$ and $U$. According to Equations 14 and 17, the temperature difference between 150 K and 300 K corresponds to a velocity ratio of only ~ $(720 / 580)^{0.5} \approx 1.1$, that is to say, the $U$ distribution of impactors for a metallic body has a more significant effect on $\frac{d}{D}$ than the temperature of the body. This is one reason why the effect of temperature difference on crater shape was not apparent (Libourel, et al., 2019).

To test the effect of impact conditions on the $\frac{d}{D}$ values of craters on the surface of an iron meteorite parent body, we calculated $\frac{d}{D}$ based on two different environments. Model 1 was for a body in the main asteroid belt and model 2 was for a body in the terrestrial planet region. In model 1, the impactor velocity distribution was assumed to mimic the population of impactors in the main asteroid belt (Bottke, et al., 1994). We used a synthesized probability function:

$$P(U)dU \propto \exp\left\{-\frac{(U-U_{01})^2}{4\sigma_1^2}\right\} \exp\left\{-\frac{(log_{10}U - log_{10}U_{01})^2}{4(log_{10}\sigma_1)^2}\right\} dU, \text{(model 1)} \quad (19a)$$

where we chose $U_{01}$ = 4.4 km s$^{-1}$ and $\sigma_1$ = 1.8 km s$^{-1}$. The typical value of the impact velocities $U$ of the bodies striking one another in the regions of 0.5–1.0 au and 1.0–1.5 au from the Sun were estimated to be ≈12 and 10 km s$^{-1}$, respectively (Bottke et al., 2006). For model 2, we chose a median $U$ of 11 km s$^{-1}$ and used the following simpler probability function:

$$P(U)dU \propto \exp\left\{-\frac{(U-U_{02})^2}{2\sigma_2^2}\right\} dU, \text{(model 2)} \quad (19b)$$

where $U_{02}$ = 11 km s$^{-1}$ and $\sigma_2$ = 3.2 km s$^{-1}$. The temperatures of the bodies were assumed to be 150 K and 300 K, corresponding to $Y_s$ = 720 MPa and 580 MPa, for models 1 and 2, respectively. The $\frac{d}{D}$ distribution was calculated using Equations 11, 18, 19, and the probability of oblique impact, $P(\theta)d\theta = 2sin\theta cos\theta d\theta$ (Shoemaker, 1962). $\rho_p$ = 3 g cm$^{-3}$ was assumed. The velocity distributions (Equations 19a and 19b) and resultant distribution of $\frac{d}{D}$ values are shown in Figure 12. The blue solid curve is the expected distribution of $\frac{d}{D}$ if cratering events on a metallic body occurred mostly in the main asteroid belt by rocky impactors after the body had cooled to the equilibrium temperature for that orbit. The apparent crater morphology may become shallower due to re-accumulated ejecta generated by impact cratering (Katsura et al., 2011). By contrast, if craters were formed by metallic impactors, the craters become deeper than the curves shown in Figure 12.

A detailed study of the surface morphology of a metallic body by a spacecraft mission, especially trying to use d/D would be a challenge, but it may reveal the past history of the impact environment of the body and will give a constraint on the dynamical evolution of the early solar system. If a metallic body migrated from the terrestrial planet region to the main asteroid belt, and if the craters on the surface of the body were formed mainly in the terrestrial planet region, then deeper craters than expected by model 1 would cover the surface of the body. If the impactor population was dominated by those of metallic, rather than rocky, composition, the $\frac{d}{D}$ values of the body's craters should be larger than in model 1.

Recent telescopic observations indicate that asteroid 16 Psyche consists of a mixture of metal and silicate, not pure iron-nickel (Shepard et al., 2017; Viikinkoski et al., 2017). Even in such a case, craters may have been formed on a surface region with a high concentration of metals or blocks of iron-rich composition. The $\frac{d}{D}$ values of micro-craters on lunar rocks provide a constraint on the composition of impacting dust particles (Hörz et al., 1975). In a similar way, the $\frac{d}{D}$ values of small craters on the surface may provide information on the composition and $U$ of impactor particles.

6. Summary

We conducted impact cratering experiments on Gibeon iron meteorite and steel targets, i.e., SS400 and iron-nickel alloy targets, at room and low temperatures, with impact velocities of 0.8–7 km s$^{-1}$. Experiments for oblique incidence using SS400 targets were also conducted and showed more prominent radial patterns downrange at room temperature than at low temperature. The crater dimensions, i.e., $D$, $d$ and $h$, were measured. We also performed iSALE-2D simulations using the JNCK model assuming JNCK parameters for the Gibeon iron meteorite, SS400 and SUS304.

The temperature dependence of $D$ was not significant, however, the $d$, and possibly $h$, values of craters in room-temperature targets were slightly larger than those in low-temperature targets. The two-dimensional numerical simulation reproduced the experimental results for Gibeon and SS400 targets. The results were compiled using non-dimensional parameters according to π-group scaling. The laboratory and numerical results collectively show that the $\frac{d}{D}$ values of metallic targets are more dependent on $U$ than are those of rocky targets. The ratio is smaller under low velocity and low-temperature conditions; however, the ratio is more sensitive to $U$ than it is to the temperature of the target. The $\frac{d}{D}$ values of craters formed by rock projectiles were shallower than those formed by metallic projectiles. Our results imply that the $\frac{d}{D}$ values of craters on metallic surfaces contain information about the past impact environment of metallic bodies.

Acknowledgments

We thank Kosuke Kurosawa, Naru Hirata, Takaya Okamoto, Masahiko Arakawa, Chisato Okamoto, and Kazunori Ogawa for fruitful discussions. We also thank Tom Davison and an anonymous reviewer for constructive and helpful reviews, and Brandon Johnson for helpful suggestions as an editor. Acknowledgements also go to Kazuyoshi Sangen for technical support for the preparation of rock projectiles and impact experiments. A. M. N. is grateful to Eileen Ryan and Don Davis for giving her a valuable opportunity to observe impact disruption experiments of cooled Gibeon targets at the NASA Ames Vertical Gas Gun Range in 1994, which inspired this work. We gratefully acknowledge the developers of iSALE-2D, including Gareth Collins, Kai Wünnemann, Dirk Elbeshausen, Tom Davison, Boris Ivanov, and Jay Melosh. This research was supported by the Hosokawa Powder Technology Foundation and by the Hypervelocity Impact Facility (former facility name: the Space Plasma Laboratory) at ISAS/JAXA.

References

Amsden, A., Ruppel, H., Hirt, C., 1980. SALE: A simplified ALE computer program for fluid flow at all speeds. Los Alamos National Laboratories Report, LA-8095:101p. Los Alamos, New Mexico: LANL.

Arai, Y., Konaka, K., Hannuki, T., Akiyama, H., 2010. Effect of the properties of Charpy impact test on steel structure bearing load. J. Struct. Constr. Eng., AIJ 75, 357-365 (in Japanese).

Benz, W., Cameron, A. G. W., Melosh, H. J., 1989. The origin of the moon and the single-impact hypothesis III. Icarus 81, 113-131.

Bottke, W. F., Nolan, M. C., Greenberg, R., Kolvoord, R. A., 1994. Velocity distributions among colliding asteroids. Icarus 107, 255–-268.

Bottke, W. F., Nesvorny, D., Grimm, R. E., Morbidelli, A., O'Brien, D.P., 2006. Iron meteorites as remnants of planetesimals formed in the terrestrial planet region. Nature 439, 821–824.

Bowling, T. J., Johnson, B. C., Melosh, H. J., Ivanov, B. A., O'Brien, D. P., Gaskell, R., Marchi, S., 2013. Antipodal terrains created by the Rheasilvia basin forming impact on asteroid 4 Vesta. J. Geophys. Res. Planets. 118, 1821-1834.

Burchell, M., J., Mackay, N. G., 1998. Crater ellipticity in hypervelocity impacts on metals. J. Geophys. Res. 103, 22,761-22,774.

Buchwald, V.F, 1969. The Gibeon meteorites. Meteoritics 4, 264-265.

Burkhardt, C., Kleine, T., Bourdon, B., Palme, H., Zipfel, J., Friedrich, J., Ebel, D., 2008. Hf–W mineral isochron for Ca, Al rich inclusions: age of the solar system and the timing of core formation in planetesimals. Geochim. Cosmochim. Acta 72, 6177–6197.

Chapman C. R., McKinnon W. B., 1986. Cratering of planetary satellites. In: Satellites, eds. Burns J. A., Matthews M. S. Tucson, AZ, The University of Arizona Press. pp. 492–580.

Christiansen, E. L., 1992. Performance equation for advanced orbital debris shields. AIAA paper

No. 92-1462 (8 pages).

Collins G. S., Melosh H. J., Ivanov B. A., 2004. Modeling damage and deformation in impact simulations. Meteoritics & Planetary Science 39, 217–231.

Collins, G. S., Elbeshausen, D., Davison, T. M., Robbins, S. J., Hynek, B., M., 2011. The size-frequency distribution of elliptical impact craters. Earth and Planet. Sci. Lett. 310, 1-8.

Davison, T. M., Collins, G., Elbeshausen, D., Wünnemann, K., Kearsley, A., 2011. Numerical modeling of oblique hypervelocity impact on strong ductile targets. Meteoritics & Planet. Sci. 46, 1510-1524.

Elkins-Tanton, L. T., Asphaug, E., Bell, J. F., III, Bercovici, D., Bills, B. G., Binzel, R. P., Bottke, W. F., Brown, M., Goldsten, J., Jaumann, J., Jun, I., Lawrence, D. J., Lord, P., Marchi, S., McCoy, S., Oh, D., Park, R., Peplowski, P. N., Polanskey, C. A., Potter, D., Prettyman, T. H., Raymond, C. A., Russell, C. T., Scott, S., Stone, H., Sukhatme, K. G., Warner, N., Weiss, B. P., Wenkert, D. D., Wieczorek, M., Williams, D., Zuber, M. T., 2017. Asteroid (16) Psyche: Visiting a metal world. In: 48$^{th}$ Lunar and Planetary Sci. Conf., the Woodlands, TX, March 20-24, 2017.

Furnish, M., Gray, G. T., Remo, J., 1994. Dynamical behavior of octahedrite from the henbury meteorite. AIRAPT/Am. Phys. Soc. Conf., June 28-July 2, Colorado Springs, Colo. AIP Conference Proceedings, Volume 309, Issue 1, p.819-822.

Ganino, C., Libourel, G., Nakamura, A. M., Michel, P., 2019. Are hypervelocity impacts able to produce chondrule-like ejecta? Planet. Space Sci. 177, 104684, doi: 10.1016/j.pss.2019.06.008.

Gault, D. E., 1973. Displaced mass, depth, diameter, and effects of oblique trajectories for impact craters formed in dense crystalline rocks, Moon, 6, 32–44, doi:10.1007/BF02630651.

Goldberg, E., A. Uchiyama, It. Brown, 1951. The distribution of nickel, cobalt, gallium, palladium and gold in iron meteorites, Geochem. Cosmochim. Acta, 2, 1-25.

Gordon, R. B., 1970. Mechanical properties of iron meteorites and the structure of their parent planets. J. Geophys. Res. 75, 439–447.

Haack, H., McCoy, T. J., 2003. Iron and stony-iron meteorites. In: Meteorites, Comets and Planets, ed. A. M. Davis, Vol. 1 Treatise on Geochemistry, edited by Holland, H. D. and Turekian, K. K., 325–345. Elsevier-Pergamon.

Hörz, F., Brownlee, D. E., Fechtig, H., Hartung, J. B., Morrison, D. A., Neukum, G., Schneider, E., Vedder, J. F., Gault, D. E., 1975. Lunar microcraters: Implications for the micrometeoroid complex. Planet. Space Sci. 23, 151-172.

Holsapple, K, Schmidt., R., 1982. On the Scaling of Crater Dimensions 2. Impact Processes. J. Geophys. Res. 87, 1849-1870.

Holsapple, K, 1993. The scaling of impact processes in planetary sciences. Ann. Rev. Earth Planet. Sci., 21, 333–373.

Housen, K., Holsapple, K., 2011. Ejecta from impact craters. Icarus 211, 856-875.

Itoh, Y., Usami, K., Kusama, R. Kainuma, S., 2001. Numerical Analyses of Steel and Aluminum Alloy Bridge Guard Fences. The Eighth East Asia Pacific Conference on Structural Engineering and Construction, Singapore, December 5-7, 2001, 1332 (http://hdl.handle.net/2237/5317).

Ivanov, B. A., Deniem, D., Neukum, G., 1997. Implementation of dynamic strength models into 2D hydrocodes: Applications for atmospheric breakup and impact cratering. International Journal of Impact Engineering 20, 411-430.

Johnson, A.A., Remo, J. L., 1974. A new interpretation of the mechanical properties of the Gibeon meteorite. J. Geophys. Res. 79, 1142–1146.

Johnson, G. R., Cook, W. H., 1983. A constitutive model and data for metals subjected to large strains, high strain rates and high temperatures. In: Proc. 7th Int. Symp. on Ballistiques, pp. 541-547. The Hague, The Netherlands, April, 1983.

Johnson, G. R., Cook, W. H., 1985. Fracture characteristics of three metals subjected to various strains, strain rates, temperatures and pressures. Eng. Fract. Mech. 21, 31–48.

Katsura, T., Nakamura, A. M., Suzuki, A., Hasegawa, S., 2011. Impact experiments on collisional evolution of iron regolith. In: 42nd Lunar and Planetary Sci. Conf., the Woodlands, TX, March 7-11, 2011.

Katsura, T., Nakamura, M. A,. Takabe, A., Okamoto, T., Sangen, K., Hasegawa, S., Liu, X., Mashimo, T., 2014. Laboratory experiments on the impact disruption of iron meteorites at temperature of near-Earth space. Icarus 241, 1–12.

Kawai, N., Tsurui, K., Hasegawa, S., Sato, E., 2010. Single microparticle launching method using two-stage light-gas gun for simulating hypervelocity impacts of micrometeoroids and space debris. Rev. Sci. Instrum. 81, 115105 (4 pages).

Li, X., Chen, J., YE, L., Ding, W., Song, P. 2013. Influence of Strain Rate on Tensile Characteristics of SUS304 Metastable Austenitic Stainless Steel. Acta Metall. Sin. (Engl. Lett.) 26 657-662.

Libourel, G., Nakamura, A. M., Beck, P., Potin, S., Ganino, C., Jacomet, S., Ogawa, R., Hasegawa, S., Michel, P. Hypervelocity impacts as a source of deceiving surface signatures on iron-rich asteroids. Science advances 5, eaav3971 (11 pages).

Marchi, S., Durda, D. D., Polanskey, C. A., Asphaug, E., Bottke, W. F., Elkins-Tanton, L. T., Garvie, L. A. J., Ray, S., Chocron, S., Williams, D. A., 2020. Hypervelocity impact experiments in iron-nickel ingots and iron meteorites: Implications for the NASA Psyche mission. J. Geophys. Res.: Planets 125, e2019JE005927.

Markowski, A., Leya, I., Quitté, G., Ammon, K., Halliday, A. N., Wieler, R., 2006. Correlated

helium-3 and tungsten isotopes in iron meteorites: quantitative cosmogenic corrections and planetesimal formation times. Earth Planet. Sci. Lett. 250, 104–115.

Matsui, T., Schultz, P.H., 1984. On the brittle–ductile behavior of iron meteorites: New experimental constraints. J. Geophys. Res. 89, C323–C328.

Melosh H. J., Ryan E. V., Asphaug E., 1992. Dynamic fragmentation in impacts: Hydrocode simulation of laboratory impacts. Journal of Geophysical Research 97, 14735–14759.

Nakamura, A. Fujiwara, A., 1991. Velocity distribution of fragments formed in a simulated collisional disruption. Icarus 92, 132-146.

Nakamura, A., 2017. Impact cratering on porous targets in the strength regime. Planet. Space Sci. 149, 5-13.

Nishida, M., Hayashi, K., Nakagawa, J., Ito, Y., 2012. Influence of temperature on crater and ejecta size following hypervelocity impact of aluminum alloy spheres on thick aluminum alloy targets. Int. J. Impact Eng. 42, 37-47.

Pierazzo, E., Artemieva, N., Asphaug, E., Baldwin, E. C., Cazamias, J., Coker, R., Collins, G. S., Crawford, D. A., Davison, T., Elbeshausen, D., Holsapple, K. A., Housen, K. R., Korycansky, D. G., Wünnemann, K., 2008. Validation of numerical codes for impact and explosion cratering: Impacts on strengthless and metal targets. Meteoritics & Planet. Sci. 43, 1917-1938.

Scott, E.R.D., Sanders, I.S, 2009. Implications of the carbonaceous chondrite Mn-Cr isochron for the formation of early refractory planetesimals and chondrules. Geochimica et Cosmochimica Acta, 73, 5137-5149.

Shepard, M. K., Richardson, J., Taylor, P. A., Rodriguez-Ford, L. A. Conrad, Al, de Pater, I., Adamkovics, M., de Kleer, K., Males, J. R., Morzinski, K. M., Close, L. M., Kaasalainen, M., Viikinkoski, M., Timerson, B., Reddy, V., Magri, C., Nolan, M. C., Howell, E. S.,

Benner, L. A. M., Giorgini, J. D., Warner, B. D., Harris, A. W., 2017. Radar observations and shape model of asteroid 16 Psyche. Icarus 281, 388-403.

Shimada, Y., 2012. Database of steel hysteretic characteristics affected by strain-rate. Proc. of Constructional Steel. 20, 1 (8 pages) (in Japanese).

Shoemaker, E. M., 1962. Interpretation of lunar craters. In Physics and Astronomy of the Moon. Ed. Kopal, Z. Academic Press, New York, pp. 283-359.

Suzuki, A., Hakura, S., Hamura, T., Hattori, M., Hayama, R., Ikeda, T., Kusuno, H., Kuwahara, H., Muto, Y., Nagaki, Niimi, R., Ogata, Y., Okamoto, T., Sasamori, Sekigawa, C., Yoshihara, T., Hasegawa, S., Kurosawa, K., Kadono, T., Nakamura, A. M., Sugita, S., Arakawa, M., 2012. Laboratory experiments on crater scaling-law for sedimentary rocks in the strength regime. J. Geophys. Res. 117, E08012 (7 pages)

Suzuki, A. I., Okamoto, C., Kurosawa, K., Kadono, T., Hasegawa, S., Hirai, T., 2018. Increase in cratering efficiency with target curvature in strength-controlled craters. Icarus 301, 108.

Suzuki, A. I., Fujita, Y., Harada, S., Kiuchi, M., Koumoto, Y., Matsumoto, E., Omura, T., Shigaki, E., Taguchi, S., Tsujido, S., Kurosawa, K., Hasegawa, S., Hirai, T., Tabata, M., Tamura, H., Kadono, T., Nakamura, A. M., Arakawa, M., Sugita, S., Ishibashi, K., 2021. Experimental study concerning the oblique impact of low- and high-density projectiles on sedimentary rock. Planet. Space. Sci. 195, 105141.

Swartzendruber, L. J., V. P., Itkin, C. B., Alcock, 1991. The Fe-Ni (iron-nickel) system. Jouranl of Phase Equilibria, 12, 288-312.

Thompson, S. L., Lausen, H. S, 1972. Improvements in the CHARTD radiation-hydrodynamic code III: Revised analytic equation of state. Tech. Rep. SC-RR-710714, Sandia National Laboratories, Albuquerque, NM.

Viikinkoski, M., Vernazza, P., Hanuš, J., Le Coroller, H., Tazhenova, K., Carry, B., Marsset, M.,

Drouard, A., Marchis, F., Fetick, R., Fusco, T., Ďurech, J., Birlan, M., Berthier, J., Bartczak, P., Dumas, C., Castillo-Rogez, J., Cipriani, F., Colas, F., Ferrais, M., Grice, J., Jehin, J., Jorda, L, Kaasalainen, M., Kryszczynska, A., Lamy, P., Marciniak, A., Michalowski, T., Michel, P., Pajuelo, M., Podlewska-Gaca, E., Santana-Ros, T., Tanga, P., Vachier, F., Vigan, A., Warner, B., Witasse, O., Yang, B. (16) Psyche: A mesosiderite-like asteroid? Astron. Astrophys. 619, L3 (10pp).

Walker, J. D., Chocron, S., Grosch, D. J., 2020. Size scaling of hypervelocity-impact ejecta mass and momentum enhancement: Experiments and a nonlocal-shear-band-motivated strain-rate-dependent failure model. Int. J. Impact Eng. 135, 103388.

Wasson, J. T, 1985. Meteorites: Their record of early solar system processes. New York: W. H. Freeman. 267 p.

Wünnemann, K., Collins, G., Melosh, H., 2006. A strain-based porosity model for use in hydrocode simulations of impacts and implications for transient crater growth in porous targets. Icarus, 180:514-527.

Table 1 Summary of the impact experiments using a two-stage hydrogen-gas gun

| Shot No. | projectile m (mg) | impact condition U (km/s) | impact condition θ (degree) | target T (K) | crater D (mm) | crater d (mm) | crater h (mm) |
|---|---|---|---|---|---|---|---|
| | **Dunite cylinder, 3.14ϕ ×2.0** | | | **Gibeon** | | | |
| Gd1 | 45.4 | 3.25 | 90 | room | 6.6 | 2.1 | 0.71±0.24 |
| Gd2 | 44.4 | 6.97 | 90 | room | 8.9 | 4.0 | 1.10±0.36 |
| Gd3 | 45.8 | 3.28 | 90 | 151 | 6.0 | 1.8 | 0.87±0.50 |
| | **Basalt cylinder, 3.14ϕ×2.0** | | | **Gibeon** | | | |
| Gb | 38.1 | 5.08 | 90 | 131 | 7.6 | 2.5 | 0.77±0.09 |
| | **SUS304 sphere, 1.0ϕ** | | | **Gibeon** | | | |
| Gs1 | 3.8 | 1.86 | 90 | room | 1.8 | 0.6 | 0.46±0.17 |
| Gs2 | 3.8 | 5.10 | 90 | room | 3.2 | 1.8 | 0.71±0.24 |
| Gs3 | 3.8 | 1.95 | 90 | 132 | 1.8 | 0.5 | 0.31±0.08 |
| Gs4 | 3.8 | 5.09 | 90 | 131 | 3.2 | 1.7 | 0.69±0.27 |
| | **Basalt cylinder, 3.2ϕ×2.0** | | | **SS400** | | | |
| Sb1 | 44 | 6.34 | 90 | room | 6.7 | 2.9 | 1.01±0.11 |
| Sb2 | 44 | 6.29 | 90 | ≤150[*1] | 7.7 | 3.0 | 0.90±0.14 |
| | **SUS304 sphere, 3.2ϕ** | | | **SS400** | | | |
| Ss1 | 133 | 4.89 | 90 | room | 10.6 | 6.5 | 1.61±0.44 |
| Ss2 | 133 | 4.98 | 90 | room | 10.7 | 6.1 | 1.96±0.37 |
| Ss3 | 133 | 5.01 | 90 | room | 11.3 | 6.7 | 1.79±0.49 |
| Ss4 | 133 | 5.00 | 90 | room | 11.7 | 6.9 | 1.46±0.43 |
| Ss5 | 133 | 5.10 | 90 | room | 11.6 | 7.0 | 1.34±0.41 |
| Ss6 | 133 | 4.91 | 90 | room | 11.0 | 6.8 | 1.62±0.53 |
| Ss7 | 133 | 1.98 | 90 | room | 6.7 | 2.7 | 1.26±0.05 |
| Ss8 | 133 | 5.11 | 90 | ≤150[*1] | 10.7 | 5.8 | - |
| Ss9 | 133 | 4.73 | 90 | 169 | 11.3 | 6.0 | 1.04±0.30 |
| Ss10 | 133 | 5.05 | 90 | ≤150[*1] | 11.1 | 6.3 | 1.13±0.27 |
| Ss11 | 133 | 1.88 | 90 | ≤150[*1] | 6.0 | 1.6 | 1.02±0.12 |
| SsO1 | 133 | 5.03 | 45 | room | 12.0[*2] <br> 11.0[*3] | 4.9 | - |
| SsO2 | 133 | 5.09 | 30 | room | 12.0[*2] <br> 9.1[*3] | 3.3 | - |
| SsO3 | 133 | 5.13 | 20 | room | 9.5[*2] | 2.4 | - |

| | | | | | 7.1[*3] | | |
|---|---|---|---|---|---|---|---|
| SsO4 | 133 | 5.11 | 10 | room | 9.4[*2] 3.8[*3] | 0.8 | - |
| SsO5 | 133 | 5.04 | 10 | room | 9.6[*2] 3.9[*3] | 0.8 | - |
| SsO6 | 133 | 5.00 | 45 | ≤150[*1] | 11.5[*2] 10.2[*3] | 4.3 | - |
| SsO7 | 133 | 5.05 | 30 | ≤150[*1] | 10.0[*2] 7.9[*3] | 3.0 | - |
| SsO8 | 133 | 5.07 | 20 | ≤150[*1] | 9.3[*2] 6.4[*3] | 2.0 | - |
| SsO9 | 133 | 4.98 | 10 | ≤150[*1] | 6.2[*2] 3.5[*3] | 0.7 | - |
| | **Copper sphere, 3.2ϕ** | | | **SS400** | | | |
| Sc1 | 152 | 2.38 | 90 | room | 7.5 | 3.4 | 1.28±0.34 |
| Sc2 | 152 | 4.39 | 90 | room | 11.2 | 6.3 | 1.51±0.48 |
| Sc3 | 152 | 6.52 | 90 | room | 13.4 | 8.8 | 1.90±0.51 |
| Sc4 | 152 | 4.30 | 90 | room | 11.0 | 6.2 | 1.55±0.55 |
| Sc5 | 152 | 4.35 | 90 | ≤150[*1] | 11.0 | 4.9 | 1.16±0.19 |
| Sc6 | 152 | 2.14 | 90 | ≤150[*1] | 6.7 | 2.2 | 1.34±0.33 |
| Sc7 | 152 | 5.13 | 90 | ≤150[*1] | 12.7 | 7.4 | 1.33±0.66 |
| Sc8 | 152 | 6.00 | 90 | ≤150[*1] | 13.3 | 7.8 | 1.63±0.66 |
| Sc9 | 152 | 6.08 | 90 | ≤150[*1] | 14.0 | 7.3 | - |
| | **Titanium sphere, 3.2ϕ** | | | **SS400** | | | |
| St1 | 75 | 6.50 | 90 | room | 12.1 | 5.5 | - |
| St2 | 75 | 6.48 | 90 | ≤150[*1] | 10.6 | 5.0 | 1.06±0.32 |
| | **WC-6%Co sphere, 3.2ϕ** | | | **SS400** | | | |
| Stwc1 | 250 | 1.38 | 90 | room | 4.8 | 5.9 | 1.34±0.10 |
| Stwc2 | 250 | 1.37 | 90 | ≤150[*1] | 4.5 | 4.7 | 1.25±0.07 |
| | **Aluminum sphere, 3.2ϕ** | | | **SS400** | | | |
| Sal1 | 45 | 3.49 | 90 | room | 6.9 | 2.1 | 0.78±0.18 |
| Sal2 | 45 | 3.35 | 90 | ≤150[*1] | 6.3 | 1.6 | 0.73±0.04 |
| | **Alumina sphere, 3.2ϕ** | | | **SS400** | | | |
| Sa1 | 65 | 2.23 | 90 | room | 5.7 | 2.2 | 0.84±0.09 |
| Sa2 | 65 | 2.28 | 90 | ≤150[*1] | 5.4 | 1.8 | 0.75±0.16 |

|  | **SUS304 sphere, 3.2ϕ** | | | **Iron-nickel alloy (Ni42wt%)** | | | |
|---|---|---|---|---|---|---|---|
| INs1 | 133 | 5.04 | 90 | room | 13.1 | 6.6 | 2.19±0.19 |
| INs2 | 133 | 5.08 | 90 | ≤150*[1] | 11.2 | 6.2 | 1.62±0.27 |
|  | **SUS304 sphere, 3.2ϕ** | | | **Iron-nickel alloy (Ni31.5wt%)** | | | |
| INs3 | 133 | 5.07 | 90 | room | 13.1 | 6.5 | 2.05±0.72 |
| INs4 | 133 | 5.16 | 90 | ≤150*[1] | 11.6 | 5.7 | 0.84±0.26 |

*1 The detection limit of the sensor used in the earlier shots was 150 K.

*2 Length.

*3 Width.

Table 2 Summary of impact experiments using a powder gun

| Shot No. | projectile | impact condition | | target | crater | | |
|---|---|---|---|---|---|---|---|
| | $m$ | $U$ | $\theta$ | $T$ | $D$ | $d$ | $h$*1 |
| | (g) | (km/s) | (degree) | (K) | (mm) | | |
| | **SUS304 cylinder, 15ϕ×15** | | | **SS400** | | | |
| Ss12 | 21.8 | 0.97 | 90 | room | 25.5 | 8.1 | 3.6±0.6 |
| Ss13 | 21.7 | 0.97 | 90 | room | 25.1 | 7.5 | 3.2±0.3 |
| Ss14 | 22.1 | 1.27 | 90 | room | 27.3 | 12.2 | 3.9±0.5 |
| Ss15 | 21.8 | 0.78 | 90 | 230 | 23.6 | 4.6 | 1.7±0.2 |
| Ss16 | 22.7 | 0.90 | 90 | 170 | 24.3 | 5.0 | 2.3±0.3 |
| Ss17 | 21.9 | 1.21 | 90 | 150 | 26.5 | 9.0 | 3.3±0.6 |

*1 $h$ was measured for only two points.

Table 3 Model parameters, including the Johnson-Cock (JNCK) parameters used in the iSALE 2D simulation in this study.

| Model and Parameters | | Gibeon | SUS304 | SS400 | dunite | copper |
|---|---|---|---|---|---|---|
| EOS | | ANEOS iron | | | ANEOS dunite | Tillotson |
| Strength model | | JNCK | JNCK | JNCK | ROCK | JNCK |
| Damage model | | - | - | - | IVANOV | - |
| Acoustic fluidisation | | - | - | - | BLOCK | - |
| Low density weakening | | - | - | - | POLY | - |
| Melting temperature (K) | | 1783 | 1811 | 1811 | 1373 | 1358 |
| Specific heat (J/kg K) | | 473 | 473 | 473 | 1000 | 385 |
| Thermal softening parameter | | - | - | - | 1.2 | - |
| Simon A parameter (MPa) | | 6000 | 6000 | 6000 | 1520 | 6000 |
| Simon C parameter | | 3.00 | 3.00 | 3.00 | 4.05 | 3.00 |
| Poisson's ratio | | 0.3 | 0.3 | 0.3 | 0.3 | 0.3 |
| Cohesion (undamaged, damaged) (MPa) | | - | - | - | 10, 0.01 | |
| Coefficient of internal friction (undamaged, damaged) | | - | - | - | 1.2, 0.6 | - |
| Strength at infinite pressure (GPa) | | - | - | - | 3.5 | - |
| Ivanov damage parameters | $A$ | - | - | - | $10^{-4}$ | |
| | $B$ | - | - | - | $10^{-11}$ | |
| | $C$ | - | - | - | $3 \times 10^8$ | |
| JNCK parameters | $A$ (MPa) | 676 | 468 | 360 | - | 90[*1] |
| | $B$ (MPa) | 1015 | 395 | 555 | - | 292[*1] |
| | $N$ | 0.53 | 0.54 | 0.49 | - | 0.31[*1] |
| | $C$ | 0.027 | 0.043 | 0.045 | - | 0.025[*1] |
| | $M$ | 0.52 | 0.87 | 0.60 | - | 1.09[*1] |
| | $T_{ref}$(K) | 77 | 293 | 293 | - | 293 |

[*1] Parameters of OFHC copper (Johnson and Cook, 1985).

Table 4 Results of numerical simulations.

| Shot No.[*1] | projectile mass | impact velocity | target | crater | |
|---|---|---|---|---|---|
| | $m$ | $U$ | $T$ | $D$ | $d$ |
| | (mg) | (km/s) | (K) | (mm) | |
| | **dunite cylinder, 3.15$\phi$×1.73** | | **Gibeon** | | |
| Gd1 | 44.8 | 3.25 | 298 | 6.7 | 2.0 |
| Gd2 | 44.8 | 6.97 | 298 | 9.0 | 4.0 |
| - | 44.8 | 1.00 | 132 | 3.8 | 0.5 |
| - | 44.8 | 1.50 | 132 | 4.7 | 0.7 |
| Gd3 | 44.8 | 3.28 | 132 | 6.0 | 1.8 |
| - | 44.8 | 8.00 | 132 | 9.6 | 4.4 |
| - | 44.8 | 14.0 | 132 | 12.3 | 6.0 |
| - | 44.8 | 20.0 | 132 | 14.0 | 7.4 |
| - | 44.8 | 25.0 | 132 | 15.9 | 8.9 |
| | **SUS sphere, 304 1$\phi$** | | **Gibeon** | | |
| Gs1 | 4.1 | 1.86 | 298 | 1.9 | 0.7 |
| Gs2 | 4.1 | 5.10 | 298 | 3.3 | 2.0 |
| Gs3 | 4.1 | 1.95 | 132 | 1.9 | 0.6 |
| - | 4.1 | 3.00 | 132 | 2.5 | 1.2 |
| - | 4.1 | 4.00 | 132 | 2.9 | 1.6 |
| Gs4 | 4.1 | 5.09 | 132 | 3.2 | 1.9 |
| - | 4.1 | 8.00 | 132 | 4.0 | 2.4 |
| - | 4.1 | 14.0 | 132 | 5.4 | 3.4 |
| - | 4.1 | 20.0 | 132 | 6.5 | 4.0 |
| | **Copper sphere, 3.175$\phi$** | | **SS400** | | |
| Sc1 | 150 | 2.38 | 298 | 7.8 | 4.0 |
| Sc2 | 150 | 4.39 | 298 | 10.7 | 6.5 |
| Sc3 | 150 | 6.52 | 298 | 13.1 | 8.2 |

*1 Corresponding shot number of the laboratory experiment (see Table 1).

Table 5 Fitted parameters of $\pi$-group scaling.

| Target | $K_D$ [*1] | $\alpha_D$ | $\beta_D$ | $K_d$ [*1] | $\alpha_d$ | $\beta_d$ |
|---|---|---|---|---|---|---|
| Gibeon (all data) | $10^{-0.07\pm0.30}$ | $0.27 \pm 0.15$ | $0.38 \pm 0.26$ | $10^{-1.04\pm0.20}$ | $0.55 \pm 0.09$ | $0.27 \pm 0.17$ |
| SS400 (room temp.) (SUS projectile) | $(10^{-0.062\pm0.033})$ | $0.267 \pm 0.015$ | – | $(10^{-0.786\pm0.034})$ | $0.459 \pm 0.015$ | – |
| SS400 (low temp.) (SUS projectile) | $(10^{-0.072\pm0.029})$ | $0.268 \pm 0.014$ | – | $(10^{-1.025\pm0.081})$ | $0.531 \pm 0.040$ | – |

*1 In the case of the SS400 target, the values shown are those derived with an approximation of $\pi_4^{\beta_D} = \pi_4^{\beta_d} = 1$.

Figure captions

Figure 1. Craters in Gibeon iron meteorite targets from the experiments. (a) Gd1, (b) Gd2, (c) Gd3, (d) Gb. The scale bar is 4 mm. (e) Gs1, (f) Gs2, (g) Gs3, (h) Gs4. The scale bar is 1 mm.

Figure 2. (a) The Gibeon iron meteorite sample (Gs2) before the shot. (b) A Gibeon iron meteorite sample on which we impacted a basalt projectile with a velocity of 5.0 km s$^{-1}$ at 134 K. This shot is not listed in Table 1 because we could not measure the crater dimensions.

Figure 3. Oblique impact craters on SS400 targets. (a) Incident angle, $\theta = 45°$ at room (SsO1) and (b) low temperature (SsO6), (c) $\theta = 30°$ at room (SsO2) and (d) low temperatures (SsO7), (e) $\theta = 20°$ at room (SsO3) and (f) low temperatures (SsO8), (g) $\theta = 10°$ at room (SsO4) and (h) low temperatures (SsO9), respectively. Surface elevation of the target for (i) $\theta = 20°$ at room and (j) low temperatures, (k) $\theta = 10°$ at room and (l) low temperatures. The top is downrange. The lowest elevation points of 1 or 2 pixels are not holes in the surface, but mismeasurements of the laser profiler.

Figure 4. The definitions of crater diameter, crater depth and rim height used in this study. On the left is the numerical result and on the right is the experimental result for an SUS-Gibeon 5.1-km/s

impact (Gs4). The black and gray parts of the numerical result represent the projectile and target materials, respectively.

Figure 5. The dimensions of the craters formed by the impact experiments and the numerical simulations. (a) Crater diameter versus impact energy in the experiments (except for oblique impact data), (b) crater depth versus impact energy in the experiments (except for oblique impact data), (c) depth to diameter ratios in the experiments, (d) comparisons of crater diameter, and (e) crater depth between experiments and simulations. Filled black and open blue symbols show room and low temperature data, respectively. The two data points in (c) with $d/D > 1$ are of the shots with the WC projectile. The dashed lines in (d) and (e) indicate exactly where the simulation and experiment coincide.

Figure 6. The crater dimensions for oblique impacts. (a) Ellipticity of craters. (b) Normalized average diameter and depth of craters formed by oblique impact experiments. Filled black and open blue symbols show room and low temperature data, respectively. Previous results for craters in stainless steel targets (Burchell and Mackay, 1998) are also shown. The solid curves show the power-law relationship (Equation 9) and the dashed curve shows a modified power-law relationship (Equation 10).

Figure 7. Rim height versus diameter of the craters formed at room temperature (filled black symbols) and low temperature (open blue symbols).

Figure 8. Cross-sections of the craters formed by copper projectiles (a) at room temperature with a velocity of 2.38 km s$^{-1}$ (Sc1), at low temperature (b) with velocities of 2.14 km s$^{-1}$ (Sc6), and (c) 6.08 km s$^{-1}$ (Sc9).

Figure 9. Yield strength and tensile stress of the Gibeon and Henbury meteorites. The dashed curve shows Equation 14, while the dotted curve shows the JNCK model with the parameters assumed in this study.

Figure 10. Compilation of results in π-group scaling. (a) $\pi_D$, and (b) $\pi_d$ for the laboratory data for the laboratory and numerical results in this study. Dashed curves show the fitting results for the Gibeon target.

Figure 11 Crater depth/diameter ratios for this study. The data from a previous study of a quartz projectile impacted on a Gibeon target (March et al., 2019) are also shown. Dotted curve and dashed curve show $\frac{d}{D} = 0.11\pi_3^{-0.28}$ and Equation 18, respectively.

Figure 12 (a) Velocity distribution model of asteroid main belt (model 1) and terrestrial planet region (model 2) and (b) expected crater depth/diameter ratio in a metallic body.

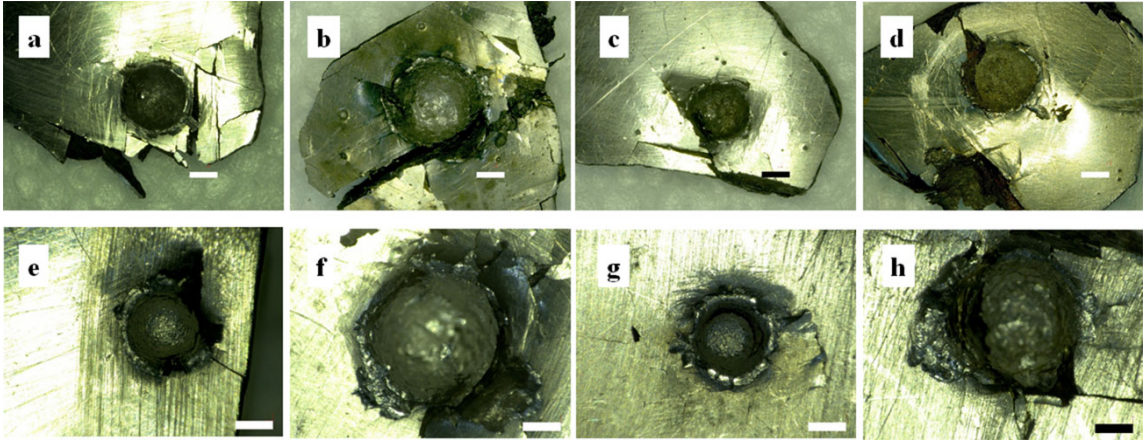

Figure 1. Craters in Gibeon iron meteorite targets from the experiments. (a) Gd1, (b) Gd2, (c) Gd3, (d) Gb. The scale bar is 4 mm. (e) Gs1, (f) Gs2, (g) Gs3, (h) Gs4. The scale bar is 1 mm.

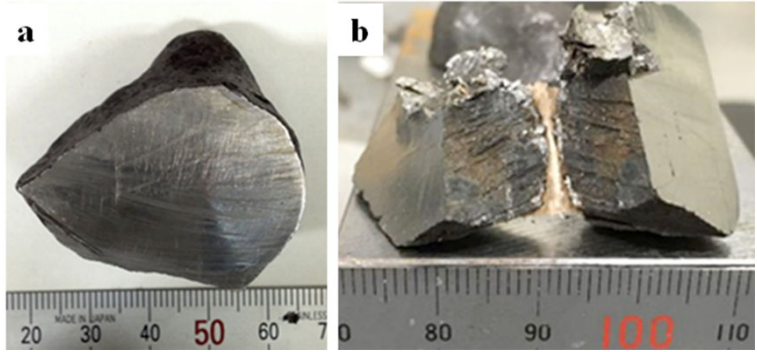

Figure 2. (a) The Gibeon iron meteorite sample (Gs2) before the shot. (b) A Gibeon iron meteorite sample on which we impacted a basalt projectile with a velocity of 5.0 km s$^{-1}$ at 134 K. This shot is not listed in Table 1 because we could not measure the crater dimensions.

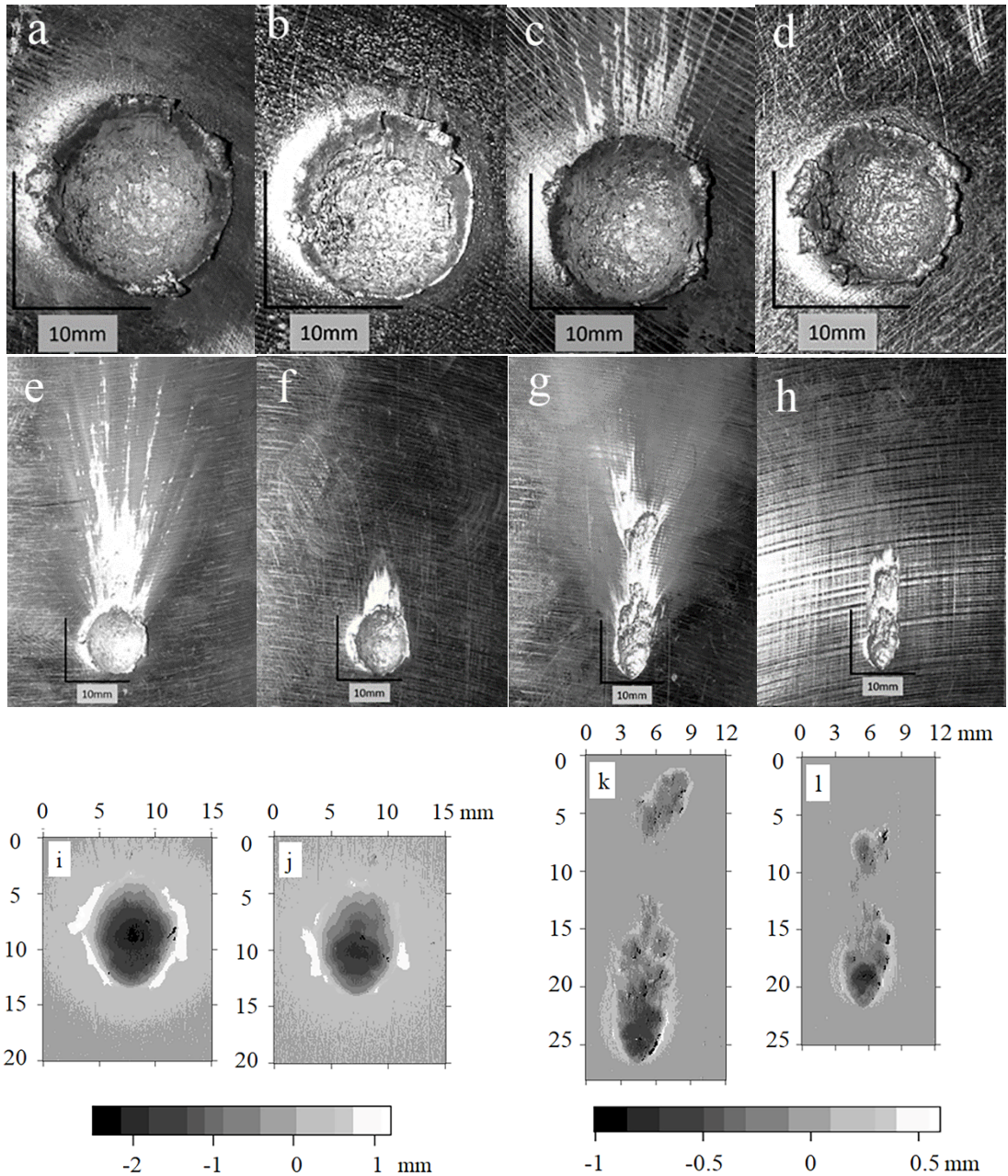

Figure 3. Oblique impact craters on SS400 targets. (a) Incident angle, $\theta = 45°$ at room (SsO1) and (b) low temperature (SsO6), (c) $\theta = 30°$ at room (SsO2) and (d) low temperatures (SsO7), (e) $\theta = 20°$ at room (SsO3) and (f) low temperatures (SsO8), (g) $\theta = 10°$ at room (SsO4) and (h) low temperatures (SsO9), respectively. Surface elevation of the target for (i) $\theta = 20°$ at room

and (j) low temperatures, (k) $\theta = 10°$ at room and (l) low temperatures. The top is downrange.

The lowest elevation points of 1 or 2 pixels are not holes in the surface, but mismeasurements of the laser profiler.

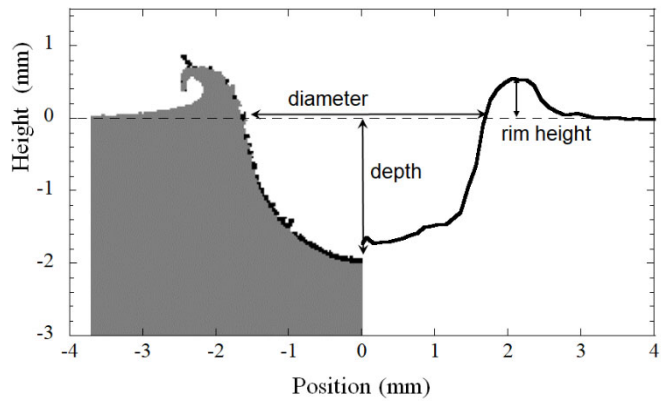

Figure 4. The definitions of crater diameter, crater depth and rim height used in this study. On the left is the numerical result and on the right is the experimental result for an SUS-Gibeon 5.1-km/s impact (Gs4). The black and gray parts of the numerical result represent the projectile and target materials, respectively.

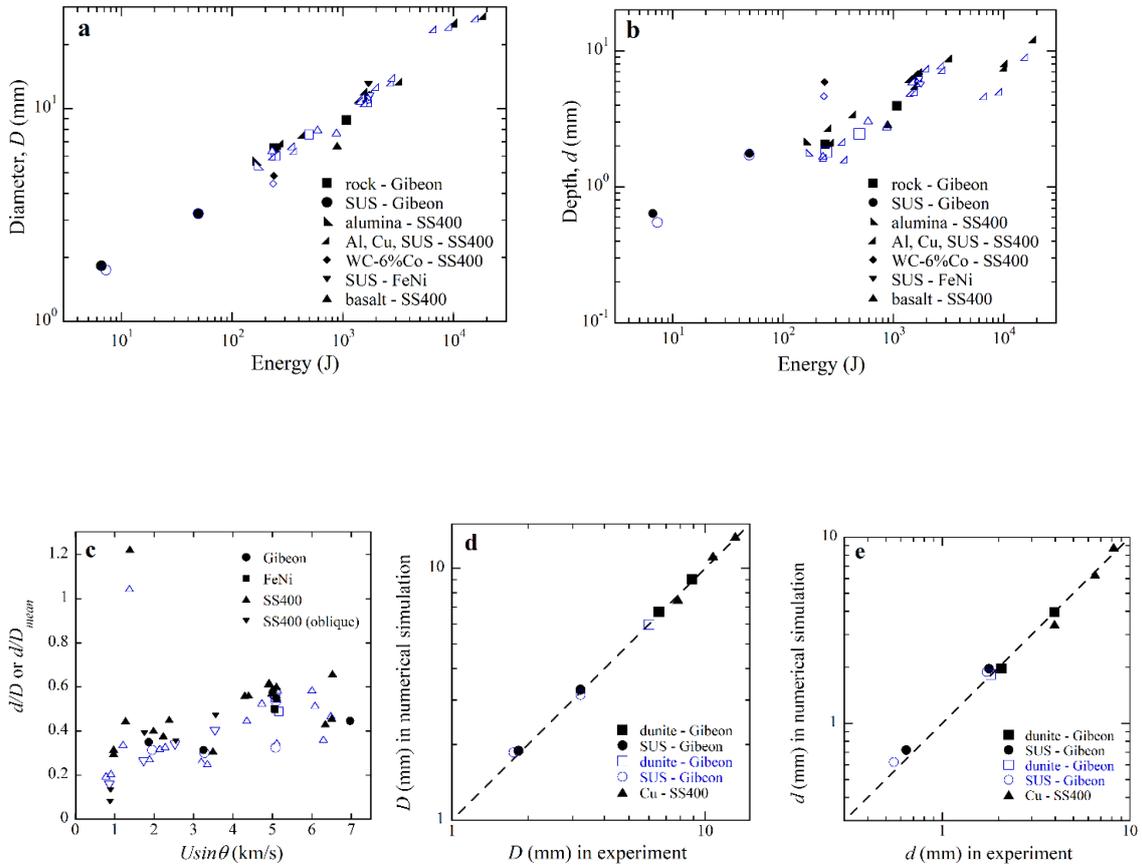

Figure 5. The dimensions of the craters formed by the impact experiments and the numerical simulations. (a) Crater diameter versus impact energy in the experiments (except for oblique impact data), (b) crater depth versus impact energy in the experiments (except for oblique impact data), (c) depth to diameter ratios in the experiments, (d) comparisons of crater diameter, and (e) crater depth between experiments and simulations. Filled black and open blue symbols show room and low temperature data, respectively. The two data points in (c) with $d/D > 1$ are of the shots with the WC projectile. The dashed lines in (d) and (e) indicate exactly where the simulation and experiment coincide.

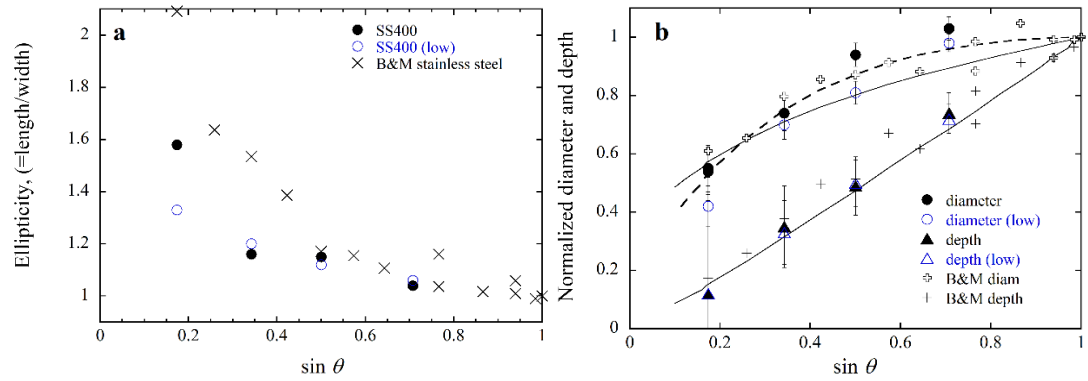

Figure 6. The crater dimensions for oblique impacts. (a) Ellipticity of craters. (b) Normalized average diameter and depth of craters formed by oblique impact experiments. Filled black and open blue symbols show room and low temperature data, respectively. Previous results for craters in stainless steel targets (Burchell and Mackay, 1998) are also shown. The solid curves show the power-law relationship (Equation 9) and the dashed curve shows a modified power-law relationship (Equation 10).

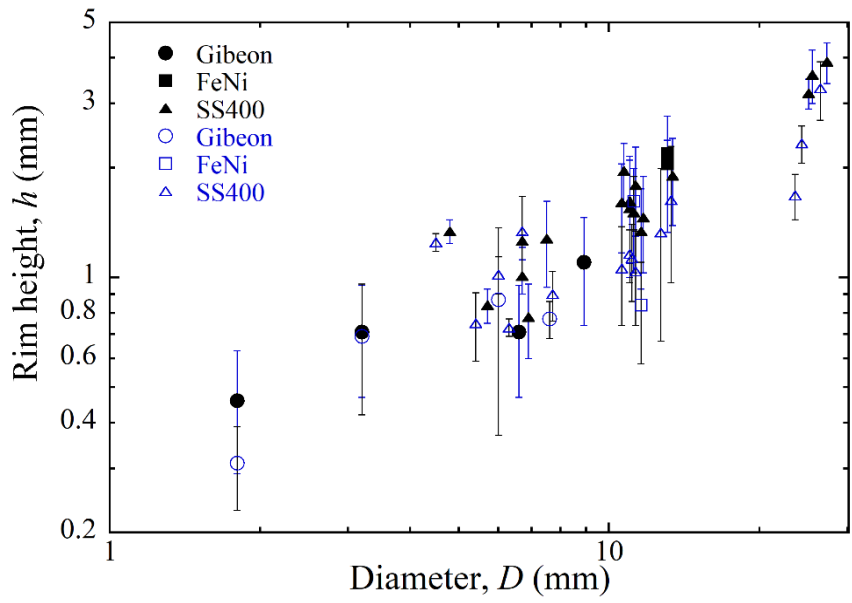

Figure 7. Rim height versus diameter of the craters formed at room temperature (filled black symbols) and low temperature (open blue symbols).

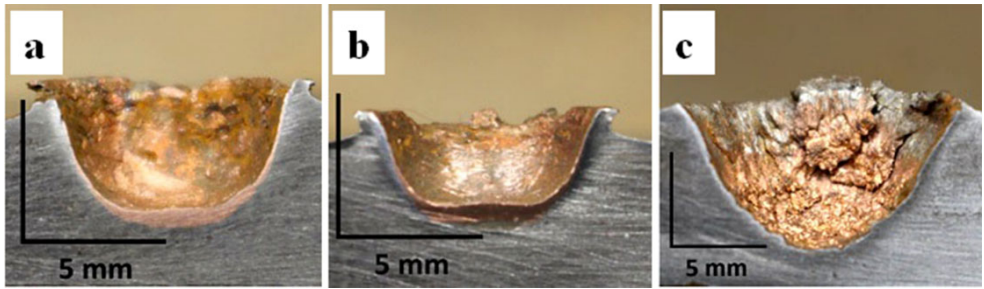

Figure 8. Cross-sections of the craters formed by copper projectiles (a) at room temperature with a velocity of 2.38 km s$^{-1}$ (Sc1), at low temperature (b) with velocities of 2.14 km s$^{-1}$ (Sc6), and (c) 6.08 km s$^{-1}$ (Sc9).

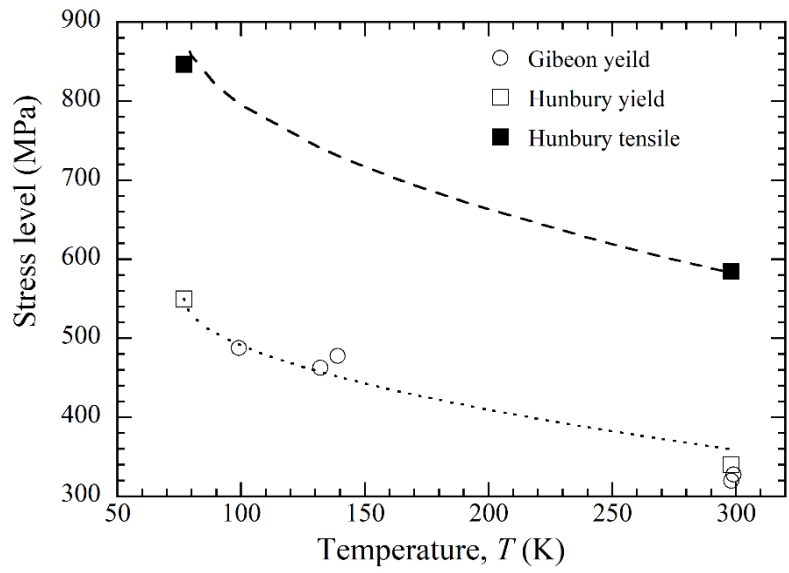

Figure 9. Yield strength and tensile stress of the Gibeon and Henbury meteorites. The dashed curve shows Equation 14, while the dotted curve shows the JNCK model with the parameters assumed in this study.

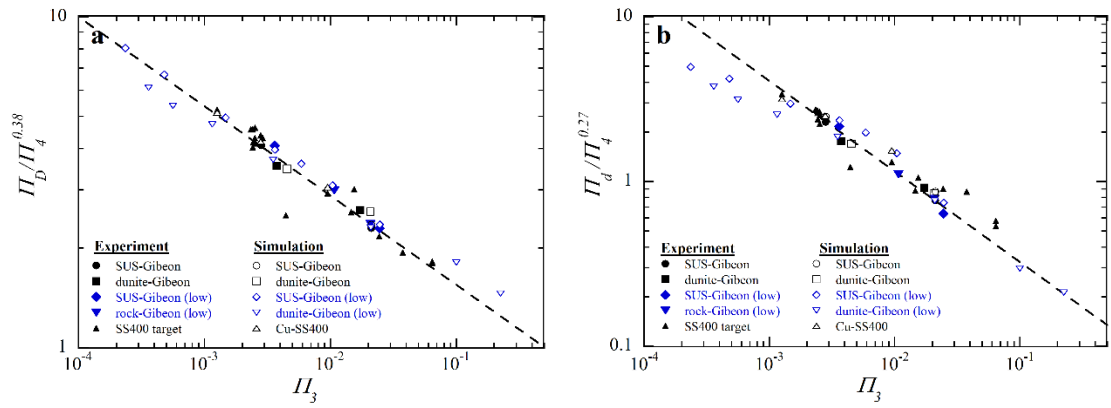

Figure 10. Compilation of results in π-group scaling. (a) $\pi_D$, and (b) $\pi_d$ for the laboratory data for the laboratory and numerical results in this study. Dashed curves show the fitting results for the Gibeon target.

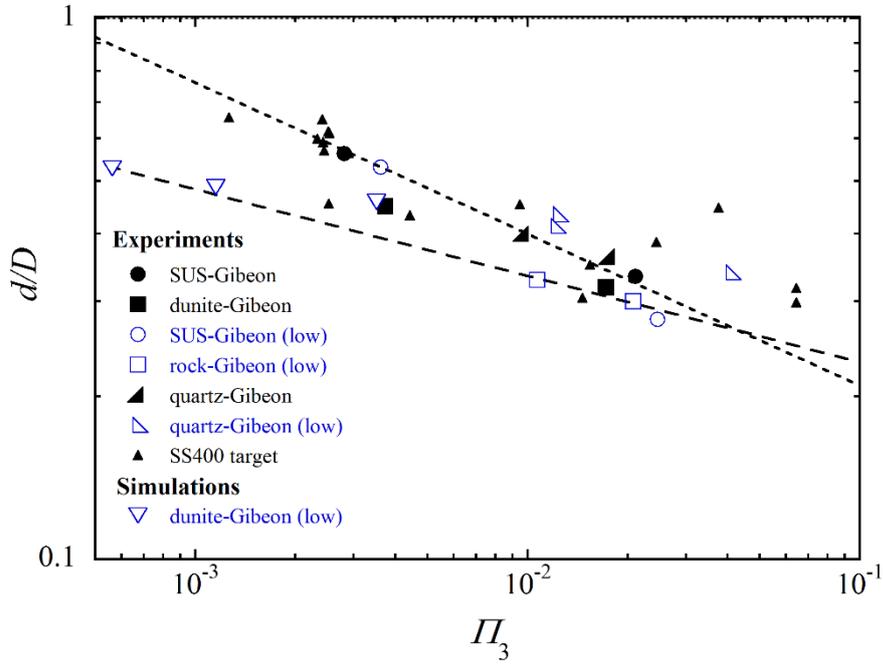

Figure 11 Crater depth/diameter ratios for this study. The data from a previous study of a quartz projectile impacted on a Gibeon target (March et al., 2019) are also shown. Dotted curve and dashed curve show $\frac{d}{D} = 0.11\pi_3^{-0.28}$ and Equation 18, respectively.

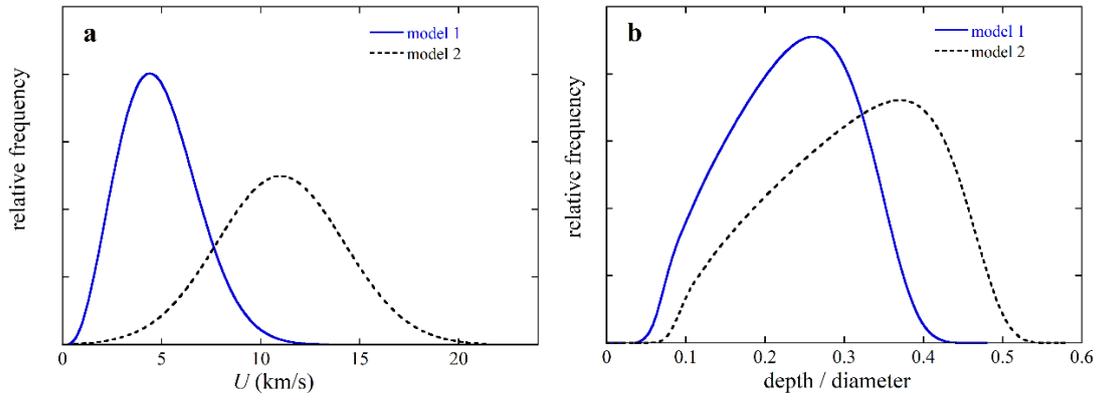

Figure 12 (a) Velocity distribution model of asteroid main belt (model 1) and terrestrial planet region (model 2) and (b) expected crater depth/diameter ratio in a metallic body.